# The Cross-Lingual Arabic Information REtrieval (CLAIRE) System

Zhizhong Chen
Brown University
Providence, RI, USA

Carsten Eickhoff
Brown University
Providence, RI, USA

## ABSTRACT

Despite advances in neural machine translation, cross-lingual retrieval tasks in which queries and documents live in different natural language spaces remain challenging. Although neural translation models may provide an intuitive approach to tackle the cross-lingual problem, their resource-consuming training and advanced model structures may complicate the overall retrieval pipeline and reduce users engagement. In this paper, we build our end-to-end Cross-Lingual Arabic Information REtrieval (CLAIRE) system based on the cross-lingual word embedding where searchers are assumed to have a passable passive understanding of Arabic and various supporting information in English is provided to aid retrieval experience. The proposed system has three major advantages: (1) The usage of English-Arabic word embedding simplifies the overall pipeline and avoids the potential mistakes caused by machine translation. (2) Our CLAIRE system can incorporate arbitrary word embedding-based neural retrieval models without structural modification. (3) Early empirical results on an Arabic news collection show promising performance.

## 1 INTRODUCTION

In the era of big data, information retrieval systems offer a convenient and automatic approach to retrieve target information for users on a large scale. Usually, queries and documents stem from the same language. This single-language assumption stands in direct conflict with the ambition of connecting users with as broad as possible a range of information. To bridge the language gap, some works rely on automatic translation of queries or documents or consider altogether multilingual users to begin with [10, 15, 19, 24]. Many multilingual users and new language learners are not equally proficient in both languages and prefer phrasing queries in the dominant language while being able to passively consume material in the second language.

In this paper, we present late-breaking results of cross-lingual query-by-example retrieval. We assume our users are English speakers with only partial proficiency in Arabic who want to explore an Arabic news wire collection. As a consequence, the majority of topic information is phrased in English, while a few limited Arabic documents are given as examples to aid model learning and to provide supporting retrieval information. To cater for this situation, we propose an end-to-end retrieval pipeline, Cross-Lingual Arabic Information REtrieval (CLAIRE) system, which 1) pre-selects potentially relevant documents via an exact-match step, 2) re-ranks them using a neural retrieval model and 3) fuses the result lists for each partial query into an overall ranking.

The remainder of this paper is structured as follows: Section 2 concisely discusses related work on cross-lingual information retrieval, query-by-example schemes, multilingual embeddings, and neural machine translation. Section 3 provides the details of the proposed CLAIRE retrieval system. Section 4 describes the experimental setup and Section 5 its results, before Section 6 closes with an outlook on future directions of inquiry.

## 2 BACKGROUND

Query-by-example (QBE) is a popular retrieval setting that offers more information to specify user's interest. It has been applied to retrieve documents or entities from the target corpus [25–27], music information retrieval [13] and tuples from relational databases [9]. In many real-world applications, the given query and the target corpus may be written in different languages. Cross-lingual information retrieval (CLIR) is essential to bridge the language gap and to ensure reliable ranking performance on this challenging task. There are two major trends: (1) translation-based approach and (2) modelling based on cross-lingual word embedding.

For translation-based approach, Douglas Oard *et al.* explore methods of translating documents into the query language [19]. And Jianfeng Gao *et al.* choose to translate queries for efficiency [10]. Also, there is a so-called interlingual approach by mapping both queries and documents into a third common language [15, 24]. Due to the effective deep representations learned from neural networks, deep learning has revolutionized the domain of machine translation significantly. Some powerful models are LSTM-Autoencoder-based Google's Neural Machine Translation system (GNMT) [30] and BERT-based translation models [6, 32]. However, training translation models can be expensive and time-consuming. Also, how to transfer pre-trained translation models to the new dataset properly and how to incorporate it into the information retrieval system effectively still remain challenging open tasks.

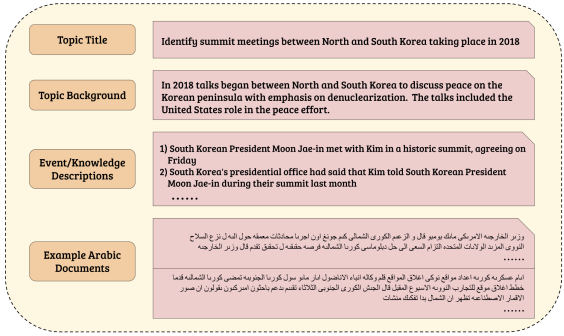

**Figure 1: Example topic.**

On the other hand, the usage of the cross-lingual word embeddings avoids the potential mistakes made by translation models. Specifically, cross-lingual word embedding maps words from different languages into the common shared vector space. It enables transferring knowledge from one language to another without the additional translation, which simplifies the overall retrieval pipeline significantly [7]. Also, it opens the retrieval system to the possibilities of exploring semantic matching signals across different languages directly and explicitly.

There are four major ways to construct cross-lingual word embeddings [23]: (1) Monolingual mapping, where a special linear mapping among various monolingual word embeddings of different languages enable them to transfer words from the source language to the target language [29]. (2) Pseudo-cross-lingual, trains word embedding models on a pseudo-cross-lingual corpus by mixing contexts of different languages [8? ]. (3) Joint optimization, jointly optimises a combination of monolingual and cross-lingual losses [5]. (4) Cross-lingual training, relies on the parallel corpus, where it builds pairs of words/sentences with similar meanings from different languages [11, 16]. In our retrieval system, we rely on the English-Arabic cross-lingual word embeddings from the work of Lachraf *et al.* [16], which is trained based on the parallel corpus.

## 3 THE CLAIRE SYSTEM

In order to simulate how non-native Arabic speakers retrieve the target Arabic information based on their knowledge in English in the realistic scenarios, we assume each topic to consist of four major components: (1) the topic title in English, which concisely describes what they are looking for, (2) the topic background in English, which gives more general supporting information, (3) related event or knowledge descriptions in English based on the user's experience, and (4) limited examples of known relevant Arabic documents. Four different components represent various knowledge aspects users obtained from different information sources. An example topic of peace efforts on the Korean peninsula is illustrated in Figure 1.

### 3.1 Retrieval Pipeline

The proposed cross-lingual query by example retrieval system, CLAIRE, comprises three major sub-systems: (1) BM25 pre-selection, (2) neural re-ranking, and (3) fusion system. The overall workflow is illustrated in Figure 2.

*BM25 Pre-selection*: We rely on BM25 [22] to filter out the majority of obviously non-relevant documents, reducing the computational burden of down-stream processing steps. To directly apply BM25 on the Arabic documents we use simple word-level English-Arabic lookup table to translate any English query components to Arabic and then apply BM25 to pre-select potentially relevant candidates.

*Neural Re-ranking*: Various neural retrieval models such as ConvKNRM [33], KNRM [3], MatchPyramid [20? ], DRMM [14] and PACRR [12] have been recently proposed. They extract semantic matching signals based on word-level embeddings of queries and documents. Usually, word embeddings are initially trained via skipgram [1] or CBOW models [17]. As a result, all word tokens of the given word embeddings are of the same language. To facilitate direct cross-lingual ranking, we instead utilise joint English-Arabic cross-lingual word embeddings [16], which is trained via the parallel English-Arabic sentence alignment. The employment of this cross-lingual embedding avoids the potential mistakes caused by translation and simplifies the overall neural retrieval re-ranking system to make it more robust and time-efficient. We follow the pre-processing and normalization approach described by Lachraf *et al.* [16] to tokenize Arabic sentences. For each component in our queries, the neural retrieval model is applied to re-rank the corresponding pre-selected Arabic documents.

*Fusion System*: Different components of our retrieval request provide partial relevance information. Rank fusion is the natural and popular technique to integrate ranking information from different sources. In the proposed CLAIRE system, Reciprocal Rank Fusion (RRF) [4] is applied to the partial ranked list produced by each topic component (title, background, description, examples). Given the ranking order of the same document $d$ in $D$ different sub-rankings, $r_1, r_2, ..., r_D$, the RRF-based ranking score in the system is given by $S_d = \sum_{k=1}^{D} 1/(r_k + 10)$.

We employ a two-step fusion due to the different ranking score calibrations between BM25 and neural retrieval models. We first apply RRF to combine ranking results of pre-selected documents for each query component based on the output of BM25 system. Then RRF is also performed on the ranking output of the neural re-ranking system. The final ranking scores in the proposed CLAIRE system are based on RRF scores of the BM25 fusion and the neural re-ranking fusion, as shown in Figure 2.

## 4 EXPERIMENTS

We conduct our retrieval experiments on a proprietary collection of 730k Arabic news articles covering a broad range of subjects. We construct 35 search topics as input to our retrieval pipeline. On average, each topic identifies 2.5 known relevant input example documents.

### 4.1 BM25 Pre-selection

The number of documents pre-selected by BM25 provides a critical foundation for our overall CLAIRE system. Too small thresholds can ignore some reasonably relevant candidates due to the simplicity of BM25. If the threshold is too large, many non-relevant documents may introduce noise or slow down the overall pipeline. Hence, we empirically investigate thresholds of 1,000, 5,000, 10,000 most highly ranked BM25 documents to be included in neural re-ranking.

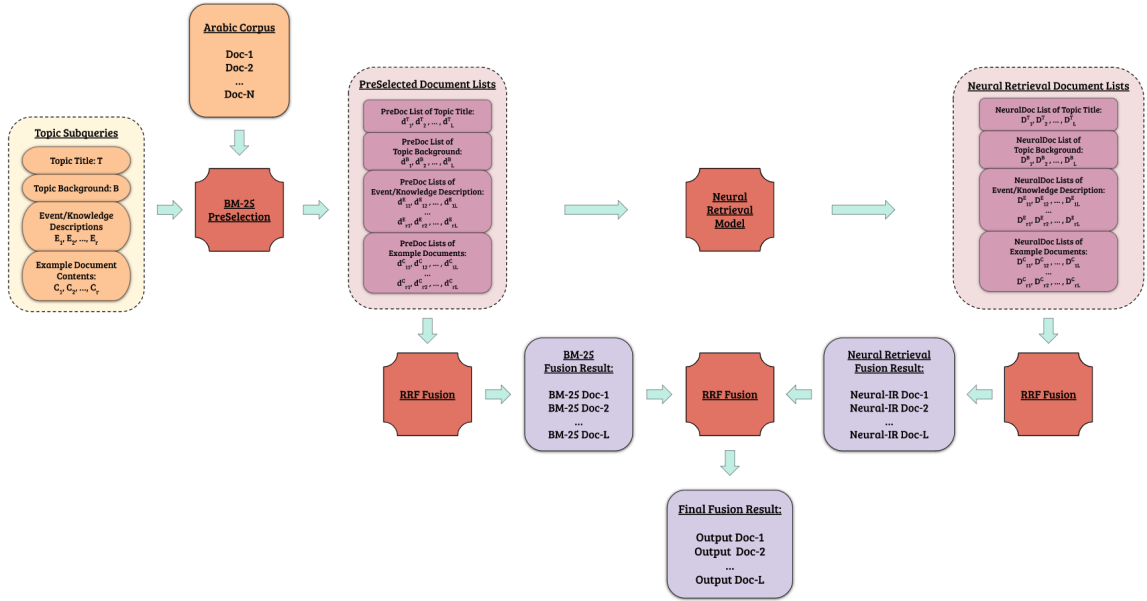

Figure 2: The overall workflow of the CLAIRE system.

## 4.2 Neural Retrieval Re-ranking

Our system trains neural retrieval models via the ListNet ranking loss [2] due to its superior ranking performance, fast convergence and scalable reliability. As a listwise approach, ListNet compares a list of candidates simultaneously and extracts the inter-document ranking signals to feed neural retrieval models. It first applies the classic softmax function to transform both candidate prediction list and relevance label list into two corresponding probability distributions, then utilises cross entropy loss to finalize optimization signals. Its simple and effective formulation often outperforms other listwise ranking schemes, such as ListMLE [31] and Approx-NDCG [21]. We choose the list size of ListNet as 50 concurrently evaluated documents. For each BM25 pre-selected candidate pool, we treat non-example documents as non-relevant and construct all possible candidate lists by randomly selecting non-example documents. Each candidate list contains only a single example document as the relevant candidate.

We consider three neural retrieval models: (1) KNRM [3], a kernel based neural retrieval model that uses Gaussian kernel pooling techniques on the word-level query-document similarity matrix to extract multi-level soft-match features. (2) ConvKNRM [33], the convolutional kernel-based extension of KNRM via Convolutional Neural Networks to compose adjacent words' n-gram embeddings and (3) MatchPyramid [20? ] that regards the constructed word-level similarity matrix from query and document embeddings as a special "image" and utilises several stacked CNN layers to extract matching signals for the final ranking score computation.

## 4.3 Model Implementation

All neural retrieval models are implemented in Tensorflow [1]. AraEngVec [16], the cross-lingual word embedding, is used for vocabulary initialization. We use an Adam optimizer to train all models for a fixed 21 epochs and save models every three epochs. Then models with the best performance are selected. The learning rate is set to 0.0001. The maximum length of queries and documents are set as 150 and 400, respectively. Regarding model-specific parameters, we apply 11 Gaussian kernels with one harvesting exact matches (mean value $\mu_0 = 1.0$ and standard deviation $\sigma_0 = 10^{-3}$) and ten capturing soft matches (mean values spaced evenly from -0.9 to 0.9, and standard deviation $\sigma_1 = 0.1$) for both KNRM and ConvKNRM. The number of CNN filters in ConvKNRM is set to 128. For MatchPyramid, we choose 3 convolution layers, each with kernel size $3 \times 3$ and 16 convolutional channels.

## 5 RESULTS

For each BM25 pre-selection threshold and neural retrieval model, we further translate the English query parts, (1) topic title, (2) topic background and (3) event/knowledge descriptions, to Arabic directly using the simple word-level English-Arabic lookup table for our neural retrieval re-ranking system as the baseline. During model evaluation, each topic query on average has around 50 relevant documents identified by a large pool of in-house expert annotators. We evaluate and compare different system settings in terms of Precision@10, Recall@10 and Normalized Discounted Cumulative Gain at the threshold of 10 (nDCG@10). Test performance is shown in Table 1.

---
[1]https://www.tensorflow.org/

Table 1: Test performance of various configurations on the news collection. EngAra indicates no translation of topic title, topic background and event/knowledge descriptions to Arabic during neural retrieval re-ranking. FullAra indicates all the query components are translated to Arabic using the simple word-level English-Arabic lookup table for neural retrieval models.

| System | | BM25 Pre-Selection | | | Neural Retrieval | | | Final Fusion | | |
|---|---|---|---|---|---|---|---|---|---|---|
| | Threshold | Precision@10 | Recall@10 | nDCG@10 | Precision@10 | Recall@10 | nDCG@10 | Precision@10 | Recall@10 | nDCG@10 |
| KNRM_EngAra | 1000 | **0.66000** | **0.16332** | **0.52567** | 0.67429 | 0.17909 | 0.51272 | 0.75429 | **0.19812** | **0.59330** |
| KNRM_FullAra | 1000 | **0.66000** | **0.16332** | **0.52567** | **0.74286** | **0.19185** | **0.54492** | **0.76000** | 0.19648 | 0.59321 |
| KNRM_EngAra | 5000 | 0.65714 | 0.16213 | 0.52439 | 0.66000 | 0.16748 | 0.48027 | 0.74857 | 0.19065 | 0.58686 |
| KNRM_FullAra | 5000 | 0.65714 | 0.16213 | 0.52439 | 0.64857 | 0.16219 | 0.46853 | 0.73143 | 0.18357 | 0.56481 |
| KNRM_EngAra | 10000 | 0.65714 | 0.16213 | 0.52408 | 0.66286 | 0.17208 | 0.48207 | 0.74571 | 0.19290 | 0.57803 |
| KNRM_FullAra | 10000 | 0.65714 | 0.16213 | 0.52408 | 0.63714 | 0.16037 | 0.45672 | 0.73143 | 0.18357 | 0.56902 |
| ConvKNRM_EngAra | 1000 | **0.66000** | **0.16332** | **0.52567** | **0.68857** | **0.18567** | **0.49862** | 0.75429 | **0.20016** | **0.59004** |
| ConvKNRM_FullAra | 1000 | **0.66000** | **0.16332** | **0.52567** | 0.67429 | 0.17642 | 0.46986 | 0.74857 | 0.19239 | 0.57362 |
| ConvKNRM_EngAra | 5000 | 0.65714 | 0.16213 | 0.52439 | 0.68571 | 0.17910 | 0.48559 | **0.75714** | 0.19755 | 0.58242 |
| ConvKNRM_FullAra | 5000 | 0.65714 | 0.16213 | 0.52439 | 0.68000 | 0.17387 | 0.49204 | 0.74571 | 0.19852 | 0.58560 |
| ConvKNRM_EngAra | 10000 | 0.65714 | 0.16213 | 0.52408 | 0.67143 | 0.16891 | 0.46591 | 0.73714 | 0.19147 | 0.57100 |
| ConvKNRM_FullAra | 10000 | 0.65714 | 0.16213 | 0.52408 | 0.66571 | 0.16711 | 0.46664 | 0.75429 | 0.19842 | 0.58043 |
| MatchPyramid_EngAra | 1000 | **0.66000** | **0.16332** | **0.52567** | **0.62857** | **0.16225** | **0.49367** | **0.72286** | **0.18736** | 0.56009 |
| MatchPyramid_FullAra | 1000 | **0.66000** | **0.16332** | **0.52567** | 0.60000 | 0.15534 | 0.47057 | 0.71429 | 0.18059 | 0.54843 |
| MatchPyramid_EngAra | 5000 | 0.65714 | 0.16213 | 0.52439 | 0.53429 | 0.13821 | 0.45464 | 0.68571 | 0.17196 | 0.54978 |
| MatchPyramid_FullAra | 5000 | 0.65714 | 0.16213 | 0.52439 | 0.47714 | 0.12244 | 0.43883 | 0.68286 | 0.17179 | 0.55628 |
| MatchPyramid_EngAra | 10000 | 0.65714 | 0.16213 | 0.52408 | 0.60571 | 0.15041 | 0.47361 | 0.71429 | 0.18066 | **0.56397** |
| MatchPyramid_FullAra | 10000 | 0.65714 | 0.16213 | 0.52408 | 0.54857 | 0.14252 | 0.45698 | 0.69429 | 0.17685 | 0.54700 |

BM25 pre-selection provides a solid foundation for the overall CLAIRE system. Due to the range of number of relevant documents for each query, a relatively low BM25 threshold of 1,000 pre-retrieved documents yields most reliable results. When there are more unnecessary documents pre-selected from BM25, the ranking performance of neural retrieval models and the final fusion results deteriorates gradually. This is especially significant for MatchPyramid model. Compared to the fully translated systems, the direct retrieval from English query components usually produces better performance in most cases, which shows the effectiveness and robustness of the joint cross-lingual word embeddings. Moreover, KNRM models provide competitive ranking performance, even better than the more sophisticated ConvKNRM in terms of nDCG@10. Since there are only 35 topics and only two or three relevant document examples for model training, simpler models can be a reasonable choice due to their weaker dependency on an abundance of training information. Furthermore, the RRF-based rank fusion improves the overall performance of the system by integrating the BM25 pre-selection system and neural retrieval re-ranking system.

## 5.1 Topic Component Analysis

To further analyse the relative importance of each topic component, we run our system for different neural retrieval models on each component individually. We again choose 1,000 as the BM25 threshold. The test performance of the final fusion is shown in Table 2. Across different neural retrieval models, the general relative importance among the four major components is consistent and clear. The Arabic example documents play an essential role and provide reliable ranking performance. Although there are no specific retrieval targets indicated in the example documents, our system can still exploit the main ideas and perform query by example retrieval effectively. On the other hand, the topic title shows relatively weaker contribution due to its limited content. This is especially difficult when there are limited amounts of data for model learning. Topic background and descriptions yield reasonable performance according to their content size. It is worth noting that even though the full fusion system performs on par with example document components alone in terms of Precision@10 and Recall@10, its nDCG@10 is significantly better.

Table 2: Topic Component Analysis on the News Collection.

| System | Final Fusion | | |
|---|---|---|---|
| | Prec@10 | Recall@10 | nDCG@10 |
| KNRM_topictitle | 0.30000 | 0.05697 | 0.23707 |
| KNRM_background | 0.60000 | 0.14053 | 0.38859 |
| KNRM_event/knwl | 0.46571 | 0.11009 | 0.39254 |
| KNRM_example | **0.76000** | **0.20199** | 0.56859 |
| KNRM_full | 0.75429 | 0.19812 | **0.59330** |
| ConvKNRM_topictitle | 0.32286 | 0.06382 | 0.24674 |
| ConvKNRM_background | 0.59714 | 0.14359 | 0.38821 |
| ConvKNRM_event/knwl | 0.49143 | 0.12517 | 0.41270 |
| ConvKNRM_example | **0.76286** | **0.20699** | 0.55419 |
| ConvKNRM_full | 0.75429 | 0.20016 | **0.59004** |
| MatchPyramid_topictitle | 0.25429 | 0.04929 | 0.21315 |
| MatchPyramid_background | 0.55714 | 0.13179 | 0.37194 |
| MatchPyramid_event/knwl | 0.40857 | 0.09662 | 0.36667 |
| MatchPyramid_example | **0.73143** | **0.19860** | 0.55538 |
| MatchPyramid_full | 0.72286 | 0.18736 | **0.56009** |

Table 3: Test Performance of Different Fusion Methods.

| Fusion Type | BM25 Pre-Selection | | | Neural Retrieval | | | Final Fusion | | |
| --- | --- | --- | --- | --- | --- | --- | --- | --- | --- |
| | Precision@10 | Recall@10 | nDCG@10 | Precision@10 | Recall@10 | nDCG@10 | Precision@10 | Recall@10 | nDCG@10 |
| KNRM_RRF | **0.66000** | **0.16332** | **0.52567** | **0.67429** | **0.17909** | **0.51272** | **0.75429** | **0.19812** | **0.59330** |
| KNRM_CombSUM | 0.61714 | 0.14313 | 0.47995 | 0.56286 | 0.15514 | 0.48261 | 0.66857 | 0.16393 | 0.52568 |
| KNRM_CombMNZ | 0.58571 | 0.13492 | 0.44652 | 0.54286 | 0.15114 | 0.47469 | 0.66857 | 0.16473 | 0.52031 |
| KNRM_ISR | 0.51429 | 0.13028 | 0.44940 | 0.62857 | 0.16093 | 0.49151 | 0.61143 | 0.15648 | 0.49245 |
| ConvKNRM_RRF | **0.66000** | **0.16332** | **0.52567** | **0.68857** | **0.18567** | **0.49862** | **0.75429** | **0.20016** | **0.59004** |
| ConvKNRM_CombSUM | 0.61714 | 0.14313 | 0.47995 | 0.62571 | 0.17019 | 0.48929 | 0.70286 | 0.17217 | 0.51779 |
| ConvKNRM_CombMNZ | 0.58571 | 0.13492 | 0.44652 | 0.62857 | 0.17064 | 0.48823 | 0.69714 | 0.17258 | 0.51527 |
| ConvKNRM_ISR | 0.51429 | 0.13028 | 0.44940 | 0.62571 | 0.16242 | 0.44688 | 0.57714 | 0.15003 | 0.48266 |
| MatchPyramid_RRF | **0.66000** | **0.16332** | **0.52567** | **0.62857** | **0.16225** | **0.49367** | **0.72286** | **0.18736** | **0.56009** |
| MatchPyramid_CombSUM | 0.61714 | 0.14313 | 0.47995 | 0.36286 | 0.09898 | 0.39256 | 0.65429 | 0.15797 | 0.51498 |
| MatchPyramid_CombMNZ | 0.58571 | 0.13492 | 0.44652 | 0.34571 | 0.09558 | 0.38472 | 0.64286 | 0.15422 | 0.51369 |
| MatchPyramid_ISR | 0.51429 | 0.13028 | 0.44940 | 0.58571 | 0.14913 | 0.47373 | 0.58000 | 0.15187 | 0.48881 |

## 5.2 Effect of the Fusion System

Besides Reciprocal Rank Fusion (RRF) [4] there are various other unsupervised rank fusion algorithms, such as CombSUM [28], CombMNZ [28] and Inverse Square Rank (ISR) [18]. Table 3 explores the effectiveness of such alternative fusion schemes. RRF-based systems outperform other fusion methods significantly across three neural retrieval models and different sub-systems. Although the difference between RRF and CombSUM is relatively small for BM25 pre-selection, CombSUM yields much worse results on neural re-ranking systems and the final fusion pipeline. In general, simple methods such as RRF and CombSUM perform better than CombMNZ and the more involved ISR. Due to the unsupervised nature of these rank fusion methods, simpler methods with fewer *a priori* assumptions are more robust and generalizable to diverse scenarios.

## 6 CONCLUSION

This chapter describes our late-breaking investigation results of cross-lingual query by example retrieval models. Our end-to-end Cross-Lingual Arabic Information REtrieval (CLAIRE) system yields promising performance without incorporating the complicated neural translation models into our neural retrieval system. To simulate how non-Arabic speakers conduct Arabic retrieval tasks in realistic scenarios, we design 35 structural topic queries to reflect various knowledge information users obtained from their experience. We use three fundamental retrieval steps: (1) BM25 pre-selection, (2) neural re-ranking, and (3) retrieval fusion. It is interesting to observe that the simpler KNRM model provides very competitive performance under the circumstances of the limited training information, compared to ConvKNRM and MatchPyramid with higher complexities. Moreover, we present the effect of different BM25 thresholds and various fusion methods. As a classic retrieval approach, BM25 provides a solid foundation of our CLAIRE system. Compared to other rank fusion methods, Reciprocal Rank Fusion (RRF) is robust and generalizable to different retrieval systems. Its reliable performance further boosts our CLAIRE system. For future work, we would like to explore more general approaches to build multi-lingual retrieval systems to aid users in exploring various languages simultaneously.

## ACKNOWLEDGMENTS

Acknowledgements blinded for anonymous review.